\documentclass[sn-mathphys-num]{sn-jnl}

\usepackage{float}
\usepackage{subcaption}
\usepackage{graphicx}%
\usepackage{multirow}%
\usepackage{amsmath,amssymb,amsfonts}%
\usepackage{amsthm}%
\usepackage{mathrsfs}%
\usepackage[title]{appendix}%
\usepackage{xcolor}%
\usepackage{textcomp}%
\usepackage{manyfoot}%
\usepackage{booktabs}%
\usepackage{algorithm}%
\usepackage{algorithmicx}%
\usepackage{algpseudocode}%
\usepackage{listings}%

\theoremstyle{thmstyleone}%
\theoremstyle{thmstyletwo}%

\theoremstyle{thmstylethree}%

\begin{document}

\title[Article Title]{Revolutionizing Quantum Mechanics: The Birth and Evolution of the Many-Worlds Interpretation}


\author*[]{\fnm{Arnub} \sur{Ghosh}}\email{arnubghosh2005@gmail.com}


\abstract{The Many-Worlds Interpretation (MWI) of quantum mechanics has captivated physicists and philosophers alike since its inception in the mid-20th century. This paper explores the historical roots, evolution, and implications of the MWI within the context of quantum theory. Beginning with an overview of early developments in quantum mechanics and the emergence of foundational interpretations, we delve into the origins of the MWI through the groundbreaking work of physicist Hugh Everett III. Everett's doctoral thesis proposed a radical solution to the measurement problem, positing the existence of multiple branching universes to account for quantum phenomena. We trace the evolution of the MWI, examining its refinement and elaboration by subsequent physicists such as John Wheeler. Furthermore, we discuss the MWI's impact on contemporary physics, including its connections to quantum information theory and ongoing experimental tests. By providing a comprehensive analysis of the MWI's historical development and current relevance, this paper offers insights into one of the most provocative interpretations of quantum mechanics and its implications for our understanding of the universe.}




\maketitle

\section{Introduction}\label{sec1}

The Many-Worlds Interpretation (MWI) of quantum mechanics proposes a startling view of reality where every quantum event branches into multiple parallel universes, each representing a different outcome [1-5]. First formulated by physicist Hugh Everett III in his doctoral thesis in 1957, the MWI challenges traditional notions of quantum measurement and the role of observers in the universe [2]. According to the MWI, every time a quantum system undergoes a measurement or interaction, the universe splits into a multitude of parallel realities, each corresponding to one of the possible outcomes. In these parallel worlds, all conceivable events and possibilities play out, resulting in a vast and diverse multiverse.
The MWI offers a solution to the long-standing conundrum known as the "measurement problem [21, 22]," which concerns the collapse of the wave function [23] and the apparent randomness of quantum events. Rather than invoking wave function collapse or hidden variables, the MWI suggests that all possible outcomes of a quantum event occur simultaneously in different branches of the multiverse. Despite its radical implications, the MWI has gained traction among physicists and philosophers for its elegance and ability to provide a straightforward explanation for quantum phenomena. However, it also raises profound questions about the nature of reality, the role of observation, and the concept of probability in quantum mechanics.

Understanding the historical development of interpretations in quantum mechanics is crucial for several reasons. Firstly, it provides insight into the intellectual journey of some of the most fundamental concepts in modern physics. By tracing the evolution of interpretations such as the Copenhagen Interpretation, Many-Worlds Interpretation, and others, researchers can appreciate the context in which these ideas emerged, the debates they sparked, and the influences they exerted on subsequent scientific thought.
Furthermore, studying the historical development of quantum mechanics interpretations offers a deeper understanding of the philosophical implications of quantum theory. It sheds light on the foundational questions surrounding the nature of reality, the role of observation, and the relationship between quantum mechanics and classical physics. Through the lens of history, researchers can explore the diverse perspectives and interpretations that have shaped our understanding of the quantum world.
Moreover, understanding the historical development of quantum mechanics interpretations enables researchers to critically evaluate current theories and interpretations. By examining the successes and shortcomings of past interpretations, scientists can identify areas for further exploration and refinement in contemporary quantum theory. Additionally, it fosters interdisciplinary dialogue between physicists, philosophers, and historians of science, enriching our collective understanding of the nature of the universe. Overall, studying the historical development of quantum mechanics interpretations is essential for advancing our knowledge of fundamental physics and addressing enduring questions about the nature of reality. 

The objectives of this paper are twofold: to explore the historical development of the Many-Worlds Interpretation (MWI) within the context of quantum mechanics and to elucidate its implications for contemporary physics and philosophy. By tracing the origins of the MWI from Hugh Everett III's seminal work to its subsequent evolution and refinement by physicists like John Wheeler, this paper aims to provide a comprehensive understanding of how this interpretation has shaped our conceptualization of quantum reality. Furthermore, it seeks to analyze the MWI's impact on modern physics, including its connections to quantum information theory, experimental tests, and philosophical implications regarding the nature of reality and probability. Through this exploration, the paper endeavors to contribute to the ongoing discourse surrounding interpretations of quantum mechanics, offering insights into one of the most intriguing and debated concepts in the field.
\section{Origins of the Many-Worlds Interpretation}\label{sec2}

\subsection*{Hugh Everett III and His Doctoral Thesis}
Hugh Everett III, born in 1930 in Washington, D.C., was destined to become one of the most influential figures in theoretical physics of the 20th century. His academic journey began at the Massachusetts Institute of Technology (MIT), where he exhibited exceptional talent and intellectual curiosity. At MIT, Everett pursued a bachelor's degree in chemical engineering, distinguishing himself as a scholar with a perfect GPA of 4.0, a testament to his exceptional aptitude and dedication to scholarly pursuits. Despite his initial focus on engineering, Everett's passion for physics soon became apparent, leading him to embark on a journey to unravel the mysteries of the universe.

In pursuit of his passion for physics, Everett enrolled in graduate studies at Princeton University, a renowned institution known for its distinguished faculty and rigorous academic standards. Under the mentorship of the esteemed physicist John Archibald Wheeler, Everett found a nurturing environment in which to cultivate his burgeoning interest in the foundations of quantum mechanics. Immersed in the rich tapestry of theoretical physics, he embarked on a quest to challenge conventional wisdom and explore the fundamental nature of reality.

During his tenure at Princeton, Everett delved into the intricacies of quantum theory, grappling with the enigmatic nature of measurement and the elusive concept of wave function collapse. His doctoral thesis, titled "Relative State Formulation of Quantum Mechanics," marked a watershed moment in the history of theoretical physics. In this seminal work, Everett dared to challenge the prevailing orthodoxy of the Copenhagen Interpretation, famously propounded by luminaries like Niels Bohr and Werner Heisenberg. With audacious clarity and intellectual rigor, he introduced the concept of quantum branching, postulating that rather than collapsing into a single outcome upon measurement, the universe bifurcates into an infinitude of parallel realities, each corresponding to a distinct outcome of the quantum event.

This revolutionary approach offered a tantalizing resolution to the age-old conundrum of measurement in quantum mechanics, providing a framework in which the superposition principle reigns supreme and determinism reigns over randomness. Despite the profound implications of his work, Everett's ideas were initially met with skepticism and resistance from the physics establishment. Undeterred by the lack of immediate acceptance, Everett remained steadfast in his convictions, eventually leaving academia to pursue a career in the defense industry.

Nevertheless, the seeds of his intellectual revolution had been sown, and over time, Everett's interpretation garnered increasing attention and respect, captivating the imaginations of subsequent generations of physicists and philosophers alike. Today, his legacy looms large in the annals of theoretical physics, serving as a beacon of inspiration for those who dare to challenge the status quo and explore the frontiers of human knowledge.

The enduring significance of Everett's contributions underscores the importance of interdisciplinary perspectives and the capacity for groundbreaking insights to emerge from unexpected sources. His work continues to stimulate debate and exploration in the realm of fundamental physics, challenging conventional wisdom and pushing the boundaries of our understanding of the quantum world. As we embark on the next chapter of scientific inquiry, Everett's visionary spirit reminds us of the boundless potential of human intellect to illuminate the mysteries of the universe.
\subsection*{Everett's Formulation of the MWI}
Everett's formulation of the Many-Worlds Interpretation (MWI) emerged as a response to the measurement problem in quantum mechanics. The measurement problem arises from the apparent collapse of the quantum wavefunction upon measurement, leading to the selection of a single outcome from a range of possible states. Hugh Everett III proposed a radical solution to this problem in his 1957 doctoral thesis, introducing the concept of quantum branching or "many worlds."
\begin{figure}[H]
\centering
\includegraphics[width=0.9\textwidth]{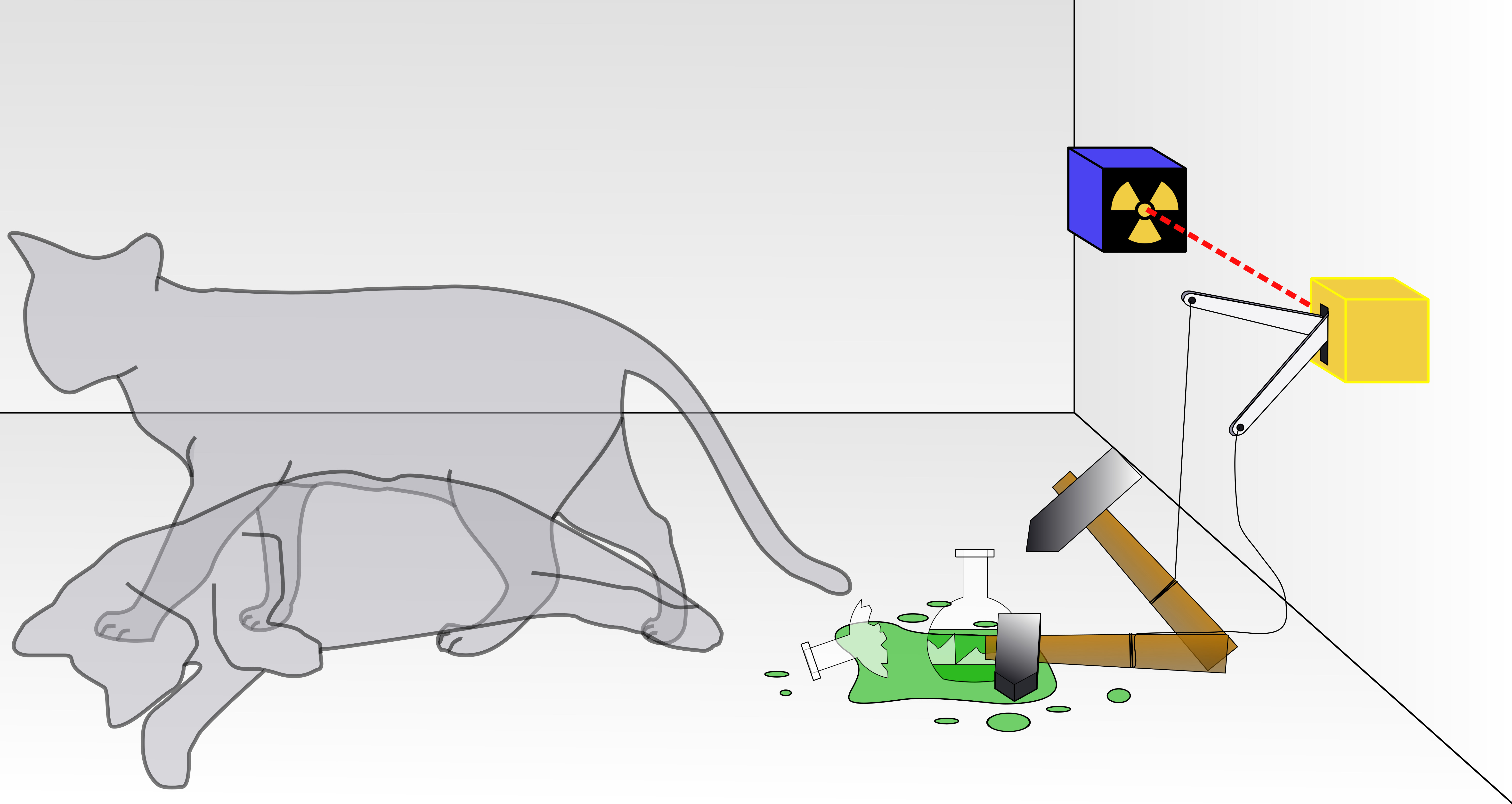}
\caption{In the Copenhagen interpretation of quantum mechanics, Schrödinger's cat is viewed as being simultaneously alive and dead until observed, illustrating the concept of superposition and the role of observation in collapsing quantum states into definite outcomes [36].}\label{fig1}
\end{figure}

In the traditional Copenhagen Interpretation, the act of measurement causes the wavefunction to collapse into one of the possible eigenstates of the observable being measured [6-10]. However, Everett argued that this collapse was merely an illusion resulting from the entanglement of the observed system with the measuring apparatus and the subsequent entanglement with the observer's consciousness. He proposed that instead of collapsing, the wavefunction of the entire universe evolves deterministically according to the Schrödinger equation.

Mathematically, the evolution of a quantum system described by the wavefunction $\psi$ can be expressed as:
\[ i\hbar \frac{\partial}{\partial t} \psi = \hat{H} \psi \]

Where $\hat{H}$ is the Hamiltonian operator representing the total energy of the system. When a measurement occurs, the system becomes entangled with the measuring apparatus, leading to the appearance of a collapse. However, Everett argued that this entanglement results in a branching of the wavefunction, rather than a collapse. 
\begin{figure}[H]
\centering
\includegraphics[width=0.9\textwidth]{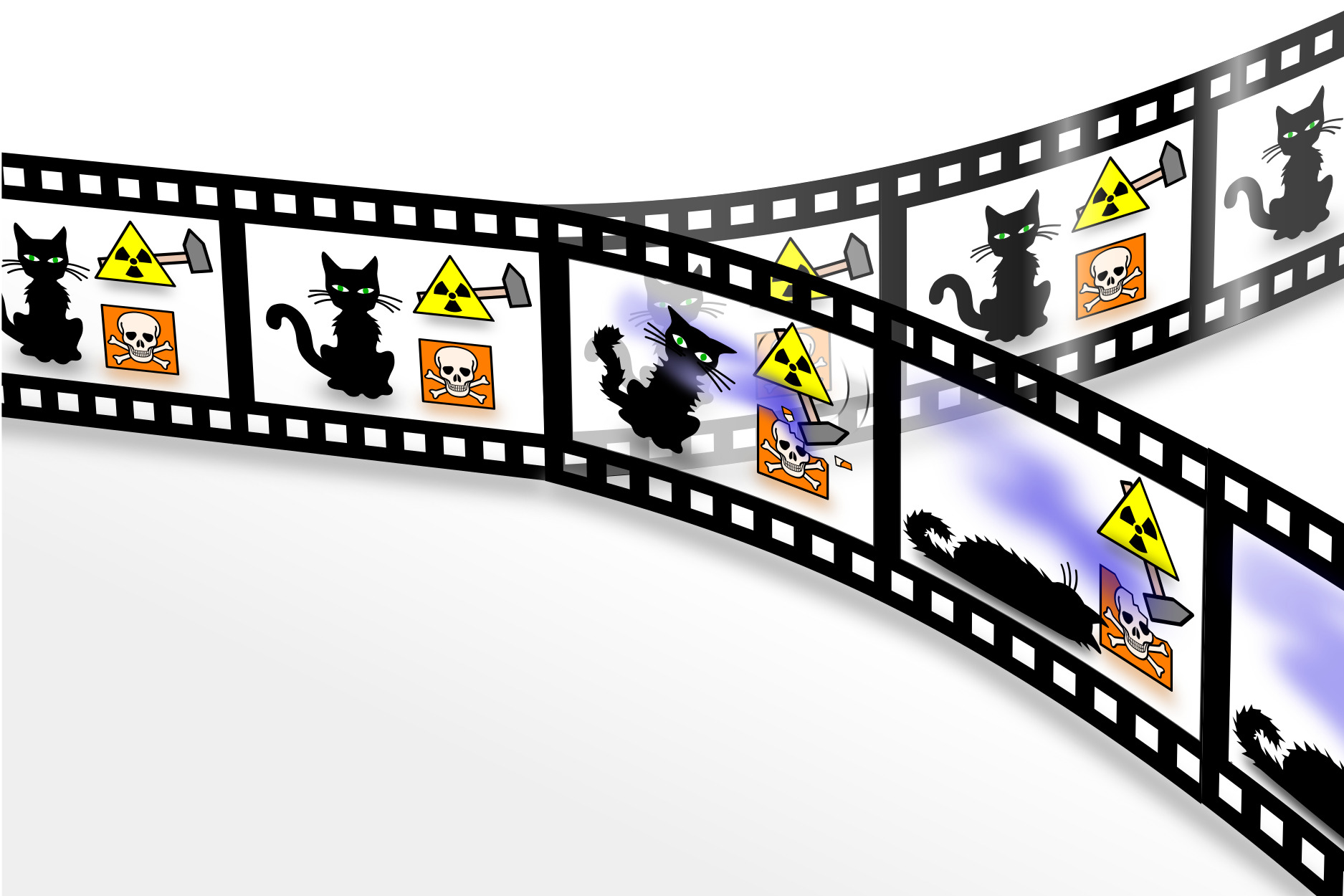}
\caption{The paradox of "Schrödinger's cat" in quantum mechanics, as viewed through the lens of the many-worlds interpretation, suggests that each quantum occurrence represents a divergence in reality. In this interpretation, the cat exists simultaneously in both living and deceased states even before the box is observed. However, these distinct states of "alive" and "dead" cats exist within separate branches of a hypothetical multiverse, each equally valid but unable to interact with one another [35].}\label{fig2}
\end{figure}

Let's consider a quantum system described by a wavefunction $\psi$, and an observable represented by an operator $\hat{A}$. The evolution of the wavefunction after measurement can be expressed as:
\[ \hat{A} \psi = \sum_{i} a_i \psi_i \]

Where $a_i$ are the possible eigenvalues of the observable, and $\psi_i$ are the corresponding eigenstates. In the MWI, instead of collapsing into one of these eigenstates, the wavefunction evolves into a superposition of all possible outcomes:
\[ \psi = \sum_{i} a_i \psi_i \]

Each term in this superposition represents a different branch of the universe, where the measurement outcome corresponds to the eigenvalue $a_i$ associated with the state $\psi_i$. Thus, rather than a single universe with a single outcome, the MWI posits the existence of a vast ensemble of parallel universes, each branching off from the others at the moment of measurement.

Everett's formulation of the MWI provided a novel and elegant solution to the measurement problem, removing the need for ad hoc postulates about the role of observers in quantum mechanics. Despite initial skepticism, the MWI has gained traction among physicists and philosophers for its simplicity and ability to account for the apparent randomness of quantum events.
\subsection*{Initial Reception and Criticisms of the Many-Worlds Interpretation (MWI)}

Hugh Everett III's groundbreaking proposal of the Many-Worlds Interpretation (MWI) in his 1957 doctoral thesis elicited a complex array of responses within the physics community, spanning from curiosity to skepticism and vigorous debate. The MWI encountered significant resistance from established figures in quantum mechanics, who were deeply entrenched in the Copenhagen Interpretation, such as Niels Bohr and Werner Heisenberg. These luminaries, along with their contemporaries, found Everett's radical departure from conventional quantum theory difficult to reconcile with their established views.

One of the primary criticisms leveled against the MWI was its apparent lack of empirical testability. Unlike other interpretations of quantum mechanics, such as the Copenhagen Interpretation or the pilot-wave theory [24], the MWI does not offer specific predictions that can be experimentally verified. This led some physicists to dismiss the MWI as unscientific or metaphysical speculation rather than a legitimate scientific theory. Additionally, the MWI faced challenges in explaining the phenomenon of quantum decoherence [16], which describes the apparent disappearance of interference effects between quantum states as a result of interactions with the environment.

Niels Bohr, a towering figure in 20th-century physics and one of the architects of the Copenhagen Interpretation, expressed skepticism towards the MWI [11]. He struggled to accept the concept of multiple parallel universes, which diverged from the central tenets of wave function collapse that formed the cornerstone of his interpretation of quantum mechanics. Similarly, Werner Heisenberg, renowned for his uncertainty principle and contributions to quantum theory, raised objections to Everett's interpretation, questioning its departure from the probabilistic nature of quantum events as described by the Copenhagen Interpretation.
\begin{figure}[H]
\centering
\begin{subfigure}[b]{0.3\textwidth}
        \includegraphics[width=\textwidth, height=6cm]{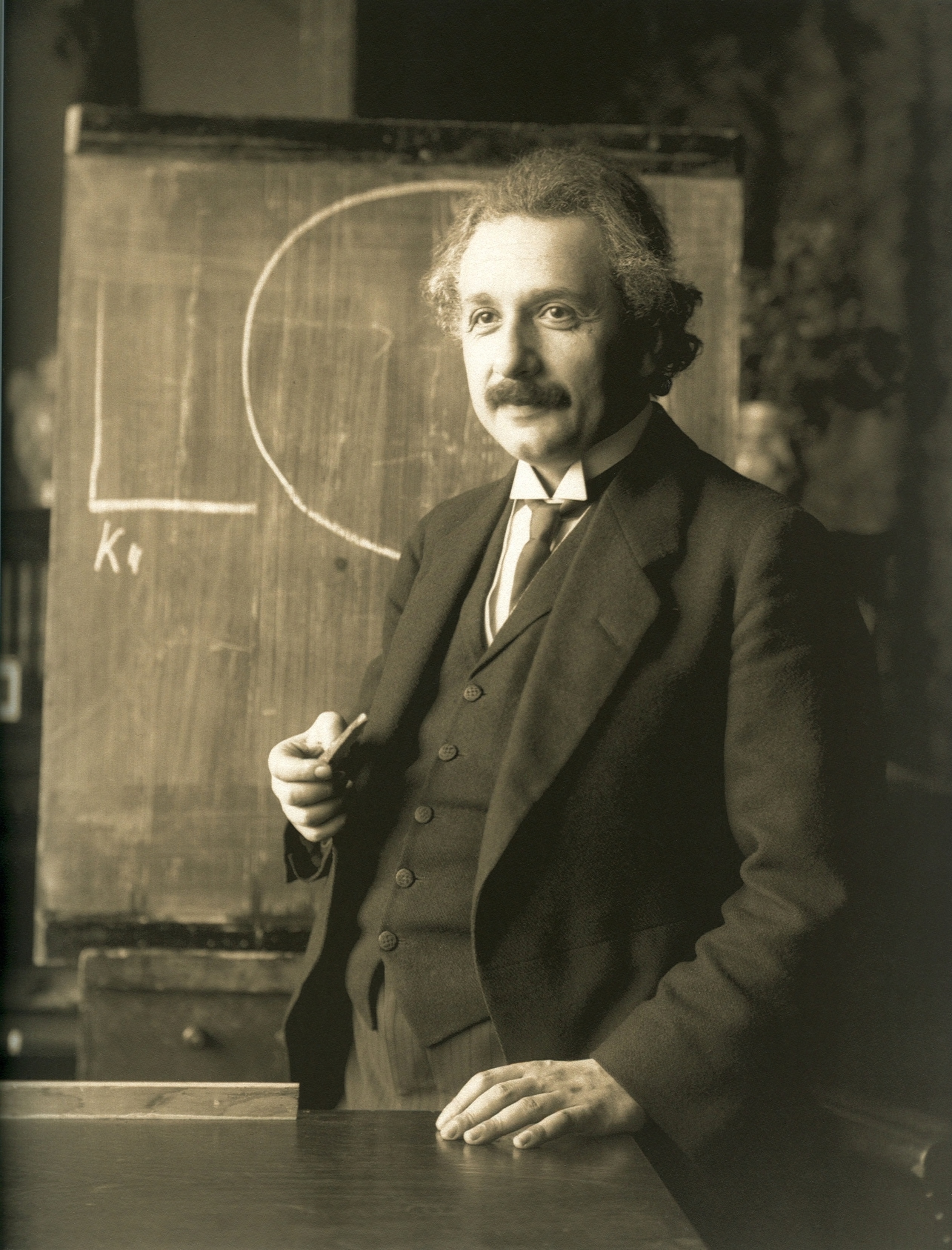}
        \caption{Albert Einstein [37]}
        \label{fig:sub1}
    \end{subfigure}
    \hfill
    \begin{subfigure}[b]{0.3\textwidth}
        \includegraphics[width=\textwidth, height=6cm]{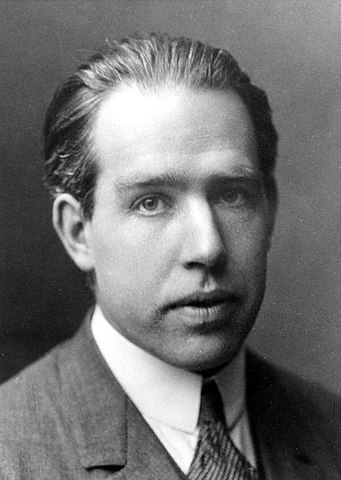}
        \caption{Niels Bohr [38]}
        \label{fig:sub2}
    \end{subfigure}
    \hfill
    \begin{subfigure}[b]{0.3\textwidth}
        \includegraphics[width=\textwidth, height=6cm]{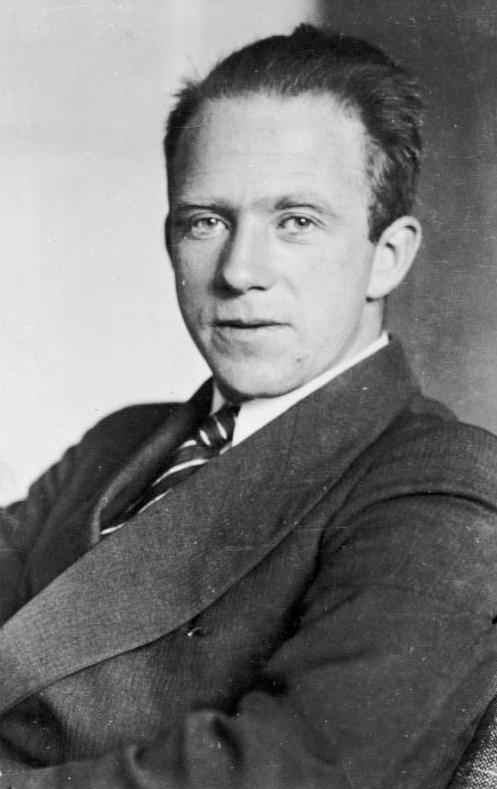}
        \caption{Werner Heisenberg [34]}
        \label{fig:sub3}
    \end{subfigure}
    \caption{Albert Einstein, Niels Bohr, and Werner Heisenberg}
    \label{fig:all_images}
\end{figure}

Eugene Wigner, another influential physicist of the era and a Nobel laureate, voiced reservations about the MWI's empirical verifiability [12]. He questioned the feasibility of experimental tests that could validate Everett's interpretation, casting doubt on its status as a scientifically viable theory.

Albert Einstein, although not directly involved in the contemporary debates surrounding the MWI, was known for his skepticism regarding the completeness of quantum mechanics [13]. Einstein's objections to the probabilistic nature of quantum theory, famously encapsulated in his assertion that "God does not play dice with the universe," resonated with critics of the MWI who were uneasy with its implications for determinism and causality.

Outside the realm of physics, philosophers also engaged with the MWI, offering critiques from philosophical and epistemological perspectives. Karl Popper, a prominent philosopher of science, questioned the empirical testability of Everett's interpretation, arguing that it lacked the falsifiability criteria essential for scientific theories [25].

Despite facing formidable criticism, the MWI gradually garnered interest and acceptance among certain segments of the physics community. Proponents of quantum information theory, such as David Deutsch and Bryce DeWitt, embraced Everett's interpretation, viewing it as a promising framework for understanding the quantum world [1, 27]. Over time, theoretical and experimental developments, including advancements in quantum computing [29] and quantum cryptography [29], provided avenues for exploring and testing the implications of the MWI.

In summary, the initial reception and subsequent criticisms of the MWI reflected the profound implications and challenges posed by Everett's radical interpretation of quantum mechanics. While skepticism and debate persisted among prominent physicists and philosophers, the MWI continued to stimulate discussion and research into the fundamental nature of reality and the interpretation of quantum phenomena.
\section{Evolution and Refinement of the Many-Worlds Interpretation
}\label{sec3}
\subsection*{Influence of John Wheeler and Other Physicists on the Development of the Many-Worlds Interpretation (MWI)}
The evolution of the Many-Worlds Interpretation (MWI) owes much to the contributions and insights of physicist John Archibald Wheeler and other luminaries who engaged with Hugh Everett III's ideas, shaping, refining, and disseminating them within the physics community [14].

\begin{figure}[H]
\centering
\includegraphics{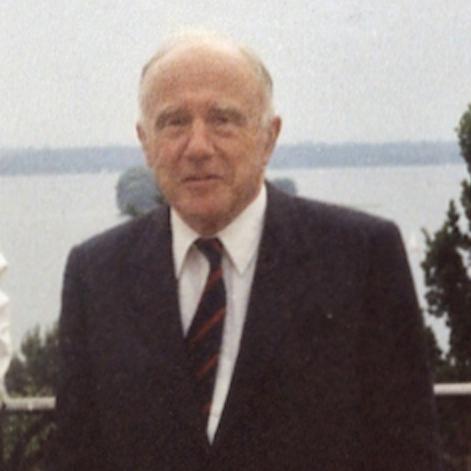}
\caption{John Archibald Wheeler in 1985 [33]}\label{fig4}
\end{figure}

John Wheeler, serving as Everett's advisor at Princeton University, played a pivotal role in the development and acceptance of the MWI. Initially skeptical of Everett's interpretation, Wheeler's stance evolved as he recognized its conceptual elegance and potential to resolve longstanding puzzles in quantum mechanics. Through his influential lectures and collaborations, Wheeler became a vocal advocate for the MWI, lending it credibility and fostering broader discussion among physicists.

Wheeler's own groundbreaking contributions to quantum mechanics and general relativity provided fertile ground for the growth of the MWI. His concept of the "participatory universe," which posited that observation fundamentally shapes reality, resonated deeply with the MWI's emphasis on the role of observers in quantum events [28]. Wheeler's openness to unconventional ideas and philosophical depth created an environment conducive to the flourishing of Everett's interpretation.

Beyond Wheeler, other physicists made significant contributions to the development and refinement of the MWI. Bryce DeWitt, a pioneer in quantum gravity, played a crucial role in popularizing the MWI through his influential papers and collaborations with Everett. DeWitt's efforts expanded the reach of the MWI within the physics community and laid the groundwork for further exploration of its implications.

David Deutsch, a proponent of quantum computation, and Lev Vaidman, a leading figure in quantum foundations, further extended Everett's ideas, exploring their implications for quantum cosmology and information theory. Their work broadened the scope of the MWI, revealing its potential applications beyond the realm of pure theory.

Moreover, the MWI garnered support from a growing cohort of researchers in quantum foundations and information science. Physicists such as Max Tegmark, Sean Carroll, and Scott Aaronson championed the MWI as a fruitful framework for understanding quantum mechanics and its implications for cosmology and consciousness. Their advocacy helped to solidify the MWI's status as a legitimate avenue of research and stimulated ongoing exploration into its ramifications.

In summary, the influence of John Wheeler and other physicists on the development of the MWI has been profound and multifaceted. Through their advocacy, insights, and collaborations, Everett's radical interpretation of quantum mechanics has gained traction and acceptance, driving continued inquiry into the nature of reality and the quantum world.

u\subsection*{Subsequent modifications and variations of the MWI}
Subsequent to its initial proposal by Hugh Everett III, the Many-Worlds Interpretation (MWI) has undergone various modifications and variations by physicists and researchers seeking to refine and expand upon its conceptual framework. These modifications reflect attempts to address perceived shortcomings, reconcile the MWI with empirical observations, and explore its implications for diverse areas of physics and philosophy.

One notable modification of the MWI is the incorporation of quantum decoherence theory, which provides a mechanism for explaining the apparent disappearance of interference effects between quantum states as a result of interactions with the environment. Physicists such as Wojciech Zurek and Maximilian Schlosshauer have developed theoretical frameworks that reconcile the MWI with decoherence theory, demonstrating how environmental interactions can lead to the emergence of classical behavior from quantum systems while preserving the integrity of multiple branches in the quantum multiverse [15].

Another significant development is the exploration of quantum Darwinism, a concept introduced by Zurek, which posits that certain observables become "redundant records" of quantum systems due to their robustness against environmental decoherence [16]. Quantum Darwinism provides a mechanism for the emergence of objective classical reality within the context of the MWI, offering insights into the origin of classical observables and their role in the quantum-to-classical transition.

Furthermore, researchers have proposed variations of the MWI that depart from Everett's original formulation in various ways. For example, the "many interacting worlds" interpretation, developed by James Hartle and Mark Srednicki, suggests that quantum coherence between different branches of the multiverse can lead to interference effects and correlations between observations made by different observers [17]. This approach emphasizes the role of interactions between branches in shaping the observed properties of the universe.

Additionally, the MWI has been extended and applied to diverse areas of physics beyond quantum mechanics, including cosmology, quantum gravity, and quantum information theory. Physicists such as Sean Carroll and Max Tegmark have explored the implications of the MWI for understanding the nature of the cosmos, the arrow of time, and the emergence of complexity in the universe [26, 30]. Moreover, researchers in quantum information science have investigated the potential advantages of the MWI for quantum computation, cryptography, and communication.

In summary, subsequent modifications and variations of the MWI reflect ongoing efforts to refine its conceptual framework, reconcile it with empirical observations, and explore its implications for diverse areas of physics and philosophy. These developments highlight the dynamism and richness of the MWI as a framework for understanding the quantum world and its relationship to the broader structure of reality.
\subsection*{Responses to critiques and ongoing debates within the physics community}
The Many-Worlds Interpretation (MWI) has elicited a range of responses and ongoing debates within the physics community, reflecting both its provocative nature and its potential implications for our understanding of quantum mechanics and reality. In response to critiques and challenges, proponents of the MWI have articulated various arguments and counterarguments, while skeptics continue to raise objections and explore alternative interpretations.

One common critique of the MWI is its apparent lack of empirical testability. Critics argue that the MWI's postulate of multiple parallel universes is inherently unobservable, making it difficult or impossible to distinguish experimentally from other interpretations of quantum mechanics. Proponents of the MWI counter that while direct observation of parallel universes may be beyond current technological capabilities, the MWI offers a coherent and mathematically elegant framework for understanding quantum phenomena and making predictions that are consistent with experimental observations. Moreover, they argue that the MWI provides a natural explanation for the apparent randomness of quantum events and resolves longstanding conceptual puzzles, such as the measurement problem.

Another criticism centers on the ontological status of the branching worlds in the MWI. Critics question whether the proliferation of parallel universes is necessary or justified, arguing that it introduces unnecessary complexity and violates principles of parsimony [32]. Proponents of the MWI respond that Occam's razor [31], which favors simpler explanations, should not be applied uncritically to the quantum realm, where intuitive notions of reality may not apply. They contend that the MWI offers a more natural and coherent interpretation of quantum mechanics than competing theories, such as the Copenhagen Interpretation or the pilot-wave theory.

Furthermore, ongoing debates within the physics community have focused on the philosophical implications of the MWI, including its implications for the nature of identity, consciousness, and free will. Some researchers argue that the MWI provides a framework for understanding the emergence of classical reality from quantum systems, shedding light on the relationship between the macroscopic world and the underlying quantum substrate. Others remain skeptical, questioning whether the MWI can fully account for the rich tapestry of human experience and subjective perception.

In summary, responses to critiques and ongoing debates within the physics community reflect the multifaceted nature of the Many-Worlds Interpretation and its implications for our understanding of quantum mechanics and reality. While proponents champion its elegance and explanatory power, skeptics raise valid concerns about its testability and philosophical implications. As research into the foundations of quantum mechanics continues, the MWI remains a fertile ground for exploration and debate, challenging physicists to grapple with the deepest mysteries of the quantum world.
\section{	MWI's impact on our understanding of quantum mechanics
}\label{sec4}
\subsection*{	MWI's impact on our understanding of quantum mechanics
}
The Many-Worlds Interpretation (MWI) has had a significant impact on our understanding of quantum mechanics, reshaping fundamental concepts and challenging conventional interpretations of quantum phenomena. By proposing that all possible outcomes of quantum events occur simultaneously in different branches of the multiverse, the MWI offers a radical reinterpretation of the quantum world and its underlying principles.

One of the key impacts of the MWI is its resolution of the measurement problem in quantum mechanics. Unlike other interpretations that invoke wavefunction collapse or hidden variables to explain the transition from quantum to classical behavior, the MWI provides a deterministic and unitary description of the evolution of the quantum state. According to the MWI, measurements are merely instances of branching within the multiverse, with each possible outcome manifesting in a separate branch. This interpretation eliminates the need for ad hoc postulates about the role of observers in the universe and provides a natural explanation for the apparent randomness of quantum events.

Furthermore, the MWI has profound implications for our understanding of quantum superposition and entanglement. In the MWI, superposition is viewed as a fundamental feature of quantum systems, with different branches of the multiverse corresponding to different states of the system. This perspective suggests that superposition is not merely a mathematical abstraction but a physical reality with tangible consequences for the behavior of quantum systems. Similarly, entanglement is interpreted as a form of correlation between branches of the multiverse, allowing for non-local connections between distant particles and phenomena.

Moreover, the MWI has led to new insights into the nature of probability in quantum mechanics. In the MWI, probabilities arise from the relative frequencies of different outcomes across multiple branches of the multiverse, rather than from inherent randomness or observer-dependent collapses of the wavefunction. This perspective offers a coherent and objective interpretation of probability in quantum mechanics, resolving longstanding debates about the nature of quantum uncertainty.

Additionally, the MWI has stimulated research into the foundations of quantum mechanics and the implications of quantum phenomena for other areas of physics. Researchers have explored the connections between the MWI and quantum information theory, quantum cosmology, and quantum gravity, seeking to understand how the principles of the MWI apply to different physical contexts. Furthermore, the MWI has inspired philosophical debates about the nature of reality, consciousness, and the relationship between the macroscopic world of classical reality and the underlying quantum substrate.

In summary, the Many-Worlds Interpretation has had a profound impact on our understanding of quantum mechanics, reshaping fundamental concepts and stimulating new avenues of research and inquiry. By providing a radical reinterpretation of quantum phenomena and their implications for the nature of reality, the MWI invites us to reconsider the fundamental principles of physics and our place within the multiverse.
\subsection*{Connections to Quantum Information Theory and Other Areas of Physics}
The Many-Worlds Interpretation (MWI) has significant connections to quantum information theory and various other areas of physics, fostering interdisciplinary research and offering new perspectives on fundamental questions in quantum mechanics.

One key connection between the MWI and quantum information theory lies in their shared emphasis on the concept of superposition. In both frameworks, superposition plays a central role in describing the state of quantum systems and the encoding of information. The MWI interprets superposition as the existence of multiple branches of the multiverse, each representing a different possible outcome of a quantum measurement. Quantum information theory, on the other hand, studies how quantum systems can be used to encode, transmit, and process information, exploiting the principles of superposition and entanglement to perform tasks such as quantum computation and cryptography. The MWI provides a natural conceptual framework for understanding the behavior of quantum information in terms of the branching structure of the multiverse, offering insights into the fundamental nature of quantum information processing.

Furthermore, the MWI has connections to quantum cosmology and the study of the universe on cosmic scales. Some researchers have proposed that the MWI may provide a natural explanation for the observed fine-tuning of the laws of physics and the existence of multiple universes within a larger "multiverse" landscape. This idea, known as the "quantum multiverse," suggests that the laws of physics may vary across different branches of the multiverse, leading to a diverse array of cosmic phenomena and possibilities. By connecting the MWI to cosmological models and observations, researchers seek to understand how the principles of quantum mechanics apply to the entire universe and its evolution over time.

Moreover, the MWI has implications for the study of quantum gravity and the unification of quantum mechanics with general relativity. Some researchers have proposed that the MWI may provide insights into the quantum nature of spacetime and the emergence of classical spacetime from underlying quantum degrees of freedom. By treating the universe as a quantum system evolving according to the principles of the MWI, researchers hope to develop a quantum theory of gravity that reconciles the discreteness of quantum mechanics with the smoothness of classical spacetime. This approach, known as quantum cosmology, seeks to understand the origin and evolution of the universe within the framework of quantum mechanics and the MWI.

In summary, the Many-Worlds Interpretation has connections to quantum information theory, cosmology, and quantum gravity, offering new perspectives on fundamental questions in physics and fostering interdisciplinary research. By connecting the principles of the MWI to other areas of physics, researchers seek to deepen our understanding of the quantum world and its implications for the nature of reality on cosmic and microscopic scales.
\subsection*{Philosophical implications}
The Many-Worlds Interpretation (MWI) of quantum mechanics carries profound philosophical implications for concepts such as reality and probability, challenging traditional views and offering new perspectives on the nature of the universe.

One of the key philosophical implications of the MWI is its redefinition of reality. In the MWI, reality is conceived as a vast and branching multiverse, encompassing all possible outcomes of quantum events. Each branch represents a distinct and equally valid reality, with every conceivable outcome of every quantum measurement manifesting in its own universe. This conception of reality challenges traditional notions of a single, objective reality and suggests that reality is inherently subjective, dependent on the perspective of the observer within the multiverse. Moreover, the MWI blurs the distinction between the macroscopic world of classical reality and the underlying quantum substrate, raising profound questions about the nature of existence and the relationship between observers and the observed.

Furthermore, the MWI has implications for our understanding of probability and randomness. In the MWI, probabilities arise from the relative frequencies of different outcomes across multiple branches of the multiverse, rather than from inherent randomness or observer-dependent collapses of the wavefunction. This perspective offers a coherent and objective interpretation of probability in quantum mechanics, resolving longstanding debates about the nature of quantum uncertainty. However, it also challenges traditional views of probability as a measure of uncertainty or ignorance, suggesting that probabilities are ontologically real and arise from the fundamental structure of the multiverse.

Moreover, the MWI has implications for our understanding of causality and determinism. In the MWI, the evolution of the multiverse is governed by the deterministic equations of quantum mechanics, with each branch representing a unique sequence of events unfolding in parallel universes. This perspective suggests that causality is an emergent property of the branching process, rather than a fundamental feature of reality. Moreover, the MWI blurs the distinction between deterministic and indeterministic interpretations of quantum mechanics, suggesting that both perspectives may be valid within the framework of the multiverse.

Additionally, the MWI raises philosophical questions about the nature of consciousness and free will. If every possible outcome of every quantum measurement occurs in a separate universe, then every possible choice or decision is realized in some branch of the multiverse. This perspective challenges traditional views of free will as the ability to choose between alternative possibilities, suggesting that free will may be illusory or relative within the context of the MWI. Moreover, the MWI raises questions about the role of observers in the universe and their ability to influence the course of events across different branches of the multiverse.

In summary, the Many-Worlds Interpretation of Quantum Mechanics has profound philosophical implications for concepts such as reality, probability, causality, consciousness, and free will. By challenging traditional views and offering new perspectives on the nature of the universe, the MWI invites us to reconsider fundamental questions about the nature of existence and our place within the multiverse.
\section{Contemporary Relevance and Future Directions
}\label{sec5}
\subsection*{Current research trends related to the MWI}
Current research trends related to the Many-Worlds Interpretation (MWI) encompass a wide range of interdisciplinary studies spanning physics, philosophy, and computational science. These trends reflect ongoing efforts to refine and expand upon the conceptual framework of the MWI, explore its implications for diverse areas of research, and develop new experimental and theoretical approaches to test its predictions.

One prominent research trend is the exploration of connections between the MWI and quantum information theory. Researchers are investigating how principles of quantum information processing, such as entanglement, quantum coherence, and quantum computation, can be understood within the framework of the MWI. This research aims to elucidate the role of information in the quantum multiverse and explore potential applications of the MWI to quantum communication, cryptography, and computation. Moreover, researchers are developing theoretical models and experimental protocols to test the predictions of the MWI in the context of quantum information processing, providing new avenues for experimental verification and validation of the MWI.

Another research trend is the investigation of connections between the MWI and quantum cosmology. Researchers are exploring how the principles of the MWI apply to the study of the universe on cosmic scales, including its origin, evolution, and large-scale structure. This research aims to develop cosmological models and observational techniques that can probe the predictions of the MWI, such as the existence of multiple universes within a larger multiverse landscape. Moreover, researchers are investigating how the MWI can shed light on fundamental questions in cosmology, such as the fine-tuning of the laws of physics and the origin of cosmic structure.

Furthermore, researchers are exploring connections between the MWI and quantum gravity, seeking to develop a unified theory that reconciles the principles of quantum mechanics with the theory of general relativity. This research aims to understand how the principles of the MWI apply to the study of spacetime, gravity, and the fundamental structure of the universe. Moreover, researchers are developing theoretical frameworks and mathematical formalisms to describe the quantum nature of spacetime within the framework of the MWI, providing new insights into the nature of black holes, wormholes, and other exotic phenomena predicted by quantum gravity theories.

In addition, researchers are exploring philosophical implications of the MWI, including its implications for concepts such as reality, probability, consciousness, and free will. This research aims to deepen our understanding of the philosophical foundations of quantum mechanics and explore the implications of the MWI for our understanding of the nature of existence. Moreover, researchers are investigating how the MWI can inform debates in philosophy of science, metaphysics, and epistemology, providing new perspectives on fundamental questions about the nature of reality and our place within the multiverse.

In summary, current research trends related to the Many-Worlds Interpretation encompass a wide range of interdisciplinary studies spanning physics, philosophy, and computational science. These trends reflect ongoing efforts to refine and expand upon the conceptual framework of the MWI, explore its implications for diverse areas of research, and develop new experimental and theoretical approaches to test its predictions.
\subsection*{Experimental tests and technological advancements}
Experimental tests and technological advancements have played a crucial role in shaping our understanding of quantum interpretations, including the Many-Worlds Interpretation (MWI) [18-20]. These endeavors have aimed to elucidate the fundamental principles of quantum mechanics, explore the predictions of different interpretations, and develop new experimental techniques to test the validity of quantum theories.

One area of experimental testing relevant to quantum interpretations involves studies of quantum superposition and entanglement. These phenomena lie at the heart of quantum mechanics and are key to distinguishing between different interpretations. Experimental tests of superposition and entanglement have provided empirical support for the predictions of quantum mechanics and have helped to rule out certain alternative interpretations that are incompatible with these phenomena. For example, experiments involving quantum interference and Bell tests have confirmed the existence of superposition and entanglement and have ruled out local hidden variable theories, which posit that quantum behavior arises from underlying deterministic processes.

Furthermore, advancements in quantum technology have enabled researchers to perform increasingly precise and sophisticated experiments to probe the foundations of quantum mechanics. Technological developments such as quantum computing, quantum cryptography, and quantum sensing have provided new tools and techniques for studying quantum phenomena and testing the predictions of different interpretations. For example, experiments using quantum computers have explored the implications of the MWI for quantum computation, demonstrating how branching in the multiverse can enhance the computational power of quantum algorithms. Similarly, experiments in quantum cryptography have tested the security of quantum communication protocols and have provided insights into the nature of quantum information processing.

Moreover, advancements in experimental techniques have enabled researchers to probe the boundary between quantum and classical physics more effectively. Experiments involving macroscopic quantum systems, such as superconducting qubits and Bose-Einstein condensates, have explored the limits of quantum coherence and the transition from quantum to classical behavior. These experiments have provided valuable insights into the mechanisms underlying decoherence and the emergence of classical reality from quantum systems, shedding light on the nature of measurement and the role of observers in quantum mechanics.

Additionally, advancements in quantum sensing and metrology have enabled researchers to study the effects of gravitational and environmental noise on quantum systems, providing new insights into the challenges of maintaining quantum coherence in practical applications. These experiments have implications for the development of quantum technologies, such as quantum computers and quantum sensors, and have highlighted the importance of understanding and mitigating decoherence effects in real-world quantum systems.

In summary, experimental tests and technological advancements have played a crucial role in shaping our understanding of quantum interpretations, including the Many-Worlds Interpretation. These endeavors have provided empirical support for the predictions of quantum mechanics, explored the implications of different interpretations, and pushed the boundaries of our knowledge of quantum phenomena. By combining theoretical insights with experimental results, researchers continue to refine and develop our understanding of the quantum world and its implications for the nature of reality.
\subsection*{Speculation on potential future developments and applications of the MWI}
Speculating on potential future developments and applications of the Many-Worlds Interpretation (MWI) entails envisioning a wide array of possibilities across multiple disciplines, ranging from physics and cosmology to philosophy and technology. While many of these potential applications remain speculative, they offer intriguing avenues for exploration and innovation.

In the realm of fundamental physics, future developments related to the MWI may involve the exploration of quantum gravity and the unification of quantum mechanics with general relativity. Researchers may seek to develop a quantum theory of gravity that incorporates the principles of the MWI, providing new insights into the nature of spacetime, black holes, and the origin of the universe. Moreover, advancements in experimental techniques and observational technologies may enable researchers to test the predictions of the MWI in the context of cosmology, probing the existence of multiple universes within a larger multiverse landscape.

Furthermore, future developments in quantum information theory may leverage the principles of the MWI to advance the field of quantum computing, cryptography, and communication. Researchers may explore how the branching structure of the multiverse can be harnessed to enhance the performance and reliability of quantum algorithms and protocols. Moreover, advancements in quantum sensing and metrology may enable the development of new technologies for precision measurements and sensing applications, leveraging the principles of superposition and entanglement to achieve unprecedented levels of sensitivity and accuracy.

In addition, future developments related to the MWI may have profound implications for our understanding of consciousness, free will, and the nature of reality. Philosophers and researchers may continue to explore the philosophical implications of the MWI, including its implications for the nature of subjective experience and the relationship between observers and the observed. Moreover, advancements in neuroscience and cognitive science may shed light on the neural correlates of conscious experience, providing new insights into the role of quantum phenomena in shaping human perception and cognition.

Moreover, future developments in technology may enable the practical realization of applications inspired by the MWI, such as quantum computing and quantum communication. Researchers may develop new quantum technologies based on the principles of the MWI, harnessing the power of superposition and entanglement to revolutionize computing, cryptography, and communication. Moreover, advancements in materials science and engineering may enable the development of new materials and devices with unique quantum properties, paving the way for a new generation of quantum technologies and applications.

In summary, speculation on potential future developments and applications of the Many-Worlds Interpretation encompasses a wide range of possibilities across multiple disciplines. While many of these potential applications remain speculative, they offer exciting opportunities for exploration and innovation, shaping our understanding of the quantum world and its implications for the nature of reality. Continued research and collaboration across disciplines will be essential to realizing the full potential of the MWI and its applications in the years to come.
\section{Conclusions}\label{sec6}
In conclusion, the exploration of the Many-Worlds Interpretation (MWI) within the historical context of quantum mechanics reveals a rich tapestry of ideas, debates, and developments that have shaped our understanding of the quantum world. Beginning with a brief overview of the MWI and its importance in the landscape of quantum interpretations, this paper delved into the historical context of quantum mechanics, tracing its origins from the early debates and challenges to the emergence of foundational interpretations such as the Copenhagen Interpretation and pilot-wave theory.
The origins of the MWI were then examined, focusing on the background of Hugh Everett III and his groundbreaking doctoral thesis, which introduced the MWI as a solution to the measurement problem in quantum mechanics. Despite initial criticisms, Everett's formulation of the MWI paved the way for subsequent modifications and refinements, influenced by physicists like John Wheeler and others who recognized its conceptual elegance and potential to address longstanding issues in quantum theory.
The implications of the MWI were explored, highlighting its impact on our understanding of quantum mechanics, its connections to quantum information theory and other areas of physics, and its philosophical implications for concepts such as reality and probability. The MWI's relevance in contemporary research trends, including experimental tests and technological advancements, was also discussed, pointing towards new avenues for exploration and innovation in the future.
In reflecting on the significance of studying the historical development of quantum interpretations, it becomes evident that the MWI represents more than just a theoretical framework—it embodies a paradigm shift in our understanding of the quantum world and its implications for the nature of reality. By tracing the evolution of the MWI and its reception within the physics community, we gain insights into the dynamic interplay between theory, experiment, and philosophy that continues to shape our understanding of the universe.
Moving forward, further research and exploration in this area hold immense potential for advancing our knowledge of quantum mechanics and its applications. By embracing the interdisciplinary nature of quantum interpretations and engaging with the historical context in which they arise, we can deepen our understanding of the fundamental principles that govern the universe and pave the way for new discoveries and insights in the years to come.

\section*{References}
\begin{enumerate}
\item Deutsch, D. (1997). The Fabric of Reality: The Science of Parallel Universes—and Its Implications. New York: Penguin Books.

\item Everett, H. III. (1957). "Relative State Formulation of Quantum Mechanics." Reviews of Modern Physics, 29(3), 454–462. https://doi.org/10.1103/RevModPhys.29.454

\item Tegmark, M. (1998). "The Interpretation of Quantum Mechanics: Many Worlds or Many Words?" In J. Pierce \& F. Greenberger (Eds.), Forty Years of Quantum Mechanics: Essays in Honor of the Celebration of the Discoveries of Max Planck (pp. 253-282). New York: Springer.

\item Wallace, D. (2012). The Emergent Multiverse: Quantum Theory According to the Everett Interpretation. Oxford: Oxford University Press.

\item Wheeler, J. A. (2000). Geons, Black Holes, and Quantum Foam: A Life in Physics. New York: W.W. Norton \& Company.

\item Bohr, N. (1935). "Can Quantum-Mechanical Description of Physical Reality Be Considered Complete?" Physical Review, 48(8), 696–702. https://doi.org/10.1103/PhysRev.48.696

\item Heisenberg, W. (1927). "Über den anschaulichen Inhalt der quantentheoretischen Kinematik und Mechanik." Zeitschrift für Physik, 43, 172–198. https://doi.org/10.1007/BF01397280

\item Mermin, N. D. (1993). "Hidden variables and the two theorems of John Bell." Reviews of Modern Physics, 65(3), 803–815. https://doi.org/10.1103/RevModPhys.65.803

\item Neumann, J. von. (1932). Mathematical Foundations of Quantum Mechanics. Princeton, NJ: Princeton University Press.

\item Peierls, R. (1991). Surprises in Theoretical Physics. Princeton, NJ: Princeton University Press.

\item Bohr, N. (1951). "Discussion with Einstein on Epistemological Problems in Atomic Physics." In P. A. Schilpp (Ed.), Albert Einstein: Philosopher-Scientist (pp. 201-241). New York: Harper \& Brothers.

\item Wigner, E. (1961). "Remarks on the Mind-Body Question." In I. J. Good (Ed.), The Scientist Speculates: An Anthology of Partly-Baked Ideas (pp. 284-302). New York: Basic Books.

\item Einstein, A. (1954). "Physics and Reality." Journal of the Franklin Institute, 257(3), 209–220.

\item Wheeler, J. A. (1957). "Assessment of Everett's 'Relative State' Formulation of Quantum Theory." Reviews of Modern Physics, 29(3), 463–465. https://doi.org/10.1103/RevModPhys.29.463

\item Zurek, W. H. (2003). "Decoherence, Einselection, and the Quantum Origins of the Classical." Reviews of Modern Physics, 75(3), 715–775. https://doi.org/10.1103/RevModPhys.75.715

\item Zurek, W. H. (2009). "Quantum Darwinism." Nature Physics, 5(3), 181–188. https://doi.org/10.1038/nphys1202

\item Hartle, J. B., \& Srednicki, M. (2007). "Are We Typical?" Physical Review D, 75(12), 123523. https://doi.org/10.1103/PhysRevD.75.123523

\item Monz, T., Schindler, P., Barreiro, J. T., Chwalla, M., Nigg, D., Coish, W. A., Harlander, M., Hänsel, W., Hennrich, M., \& Blatt, R. (2011). "14-Qubit Entanglement: Creation and Coherence." Physical Review Letters, 106(13), 130506. https://doi.org/10.1103/PhysRevLett.106.130506

\item Proietti, M., Pickston, A., Graffitti, F., Barrow, P., Kundys, D., Branciard, C., Ringbauer, M., \& Fedrizzi, A. (2019). "Experimental Rejection of Observer-Independence in the Quantum World." Science Advances, 5(9), eaaw9832. https://doi.org/10.1126/sciadv.aaw9832

\item Salart, D., Lita, A. E., Miller, A. M., \& Migdall, A. (2008). "Demonstration of Entanglement Swapping with Polarization-Entangled Photons." Physical Review Letters, 100(22), 220404. https://doi.org/10.1103/PhysRevLett.100.220404

\item Peres, A. (2002). Quantum Theory: Concepts and Methods. Dordrecht: Kluwer Academic Publishers.

\item Mermin, N. D. (1985). "Is the Moon There When Nobody Looks? Reality and the Quantum Theory." Physics Today, 38(4), 38–47. https://doi.org/10.1063/1.880968

\item Dirac, P. A. M. (1930). "A New Notation for Quantum Mechanics." Proceedings of the Cambridge Philosophical Society, 26, 376–385.

\item Bohm, D. (1952). "A Suggested Interpretation of the Quantum Theory in Terms of "Hidden" Variables. I." Physical Review, 85(2), 166–179. https://doi.org/10.1103/PhysRev.85.166

\item Popper, K. (1959). The Logic of Scientific Discovery. London: Hutchinson \& Co.

\item Carroll, S. (2010). From Eternity to Here: The Quest for the Ultimate Theory of Time. New York: Dutton.

\item DeWitt, B. (1970). "Quantum Mechanics and Reality." Physics Today, 23(9), 30–35. https://doi.org/10.1063/1.3022333

\item Wheeler, J. A. (1990). "Information, Physics, Quantum: The Search for Links." In W. H. Zurek (Ed.), Complexity, Entropy, and the Physics of Information (pp. 3-28). Redwood City, CA: Addison-Wesley.

\item Nielsen, M. A., \& Chuang, I. L. (2000). Quantum Computation and Quantum Information. Cambridge: Cambridge University Press.

\item Gisin, N., Ribordy, G., Tittel, W., \& Zbinden, H. (2002). "Quantum Cryptography." Reviews of Modern Physics, 74(1), 145–195. https://doi.org/10.1103/RevModPhys.74.145

\item Lipton, P. (2004). "Inference to the Best Explanation." Routledge Encyclopedia of Philosophy. https://doi.org/10.4324/9780415249126-S042-1.

\item Jeffrey, R. (2008). "Parsimony and Models of Learning." In E. N. Zalta (Ed.), The Stanford Encyclopedia of Philosophy. https://plato.stanford.edu/entries/parsimony-models/.

\item Emielke: \href{https://commons.wikimedia.org/w/index.php?curid=56172308}{Link} (\href{https://creativecommons.org/licenses/by-sa/3.0/}{CC BY-SA 3.0})
\item Bundesarchiv: \href{https://commons.wikimedia.org/w/index.php?curid=5436254}{Link} (\href{https://creativecommons.org/licenses/by-sa/3.0/de/}{CC BY-SA 3.0 de})
\item Christian Schirm: \href{https://commons.wikimedia.org/w/index.php?curid=17512691}{Link} (CC0)
\item Dhatfield: \href{https://commons.wikimedia.org/w/index.php?curid=4279886}{Link} (\href{https://creativecommons.org/licenses/by-sa/3.0/}{CC BY-SA 3.0})
\item Ferdinand Schmutzer/Adam Cuerden: \href{https://commons.wikimedia.org/w/index.php?curid=34239518}{Link} (Public Domain)
\item Niels Bohr's Nobel Prize biography: \href{https://commons.wikimedia.org/w/index.php?curid=288274}{Link} (Public Domain)
\end{enumerate}

\end{document}